\documentclass{elsart}
\input epsf

\newcommand{\up}{\uparrow}
\newcommand{\down}{\downarrow}

\begin{document}

\begin{frontmatter}
\title{Mechanisms of Decoherence at Low Temperatures}
\author[McGill]{M. Dub\'e} and
\author[Spinoza]{P.C.E. Stamp}

\address[McGill]{ Center for the Physics of Materials,
McGill University, 3600 rue University, Montr\a'eal, Qu\a'ebec,
Canada H3A 2T8}
\address[Spinoza]{Spinoza Institute, Universiteit Utrecht, 
Leuvenlaan 4, 3584 CE Utrecht,
The Netherlands}


\maketitle

\begin{abstract}

We briefly review the oscillator and spin bath models of 
quantum environments, which can be used to describe the 
low-energy dynamics of open quantum systems. We then use
them to discuss both the mechanisms causing decoherence at low $T$,
and the dynamics of this decoherence. This is done first 
for a central 2-level system
coupled to these environments- the results
can be applied to the dynamics of quantum nanomagnets
and superconducting SQUIDs, where large-scale tunneling of 
magnetisation and flux take place. Decoherence in these systems is caused 
principally by coupling to electrons and nuclear spins- the spin bath 
couplings are particularly dangerous at low $T$.

We may also generalise these models to discuss 
quantum measurements  
in which the measured system, the measuring apparatus, and the environment
are treated quantum mechanically. The results can be used to calculate
the dynamics of coupled SQUIDs and/or nanomagnets, in which one acts
as a measuring apparatus and the other exhibits large-scale
quantum superpositions. The same model can be used to describe coupled 
qubits.

\end{abstract}

\end{frontmatter}

\section{Introduction}

The physics of ``open''  quantum systems encompasses a vast array
of phenomena, ranging from sub-nuclear to cosmological scales.
Nevertheless, in this thematic kaleidoscope there are a few central problems, whose generality makes them relevant at almost
any scale. Here we look at one such problem --  focussing 
on low 
energy scales (ie., low-T) where clear
experimental tests of theory are possible. The problem is that of
decoherence -- where it comes from, and how to understand its effect on the 
dynamics of quantum effects involving many degrees of freedom, 
at both microscopic and macroscopic scales.

Discussions of decoherence have often been associated with rather 
ineffable questions about quantum measurements, 
quantum cosmology, and quantum computation and some of this literature 
has tended to wander around in circles, saying little that might
be quantitatively testable in the real world. This has given the 
subject an 
eccentric reputation, which is 
unfortunate, since many experiments are now 
searching for (and in some cases finding) macroscopic quantum phenomena in 
magnets and superconductors. At the same time entanglement experiments in 
optical cavities are showing multi-atom entangled states. In both kinds of 
experiment the crucial stumbling block preventing progress is 
decoherence coming from, amongst other things, coupling to environmental 
modes. The decoherence problem is thus an ``open quantum systems problem''
{\it par excellence}; it provides an extremely severe test of our 
understanding of the quantum dynamics of complex systems. Since 
the same physical mechanisms
responsible for decoherence are involved in many other 
processes of relaxation in the quantum 
regime (as well as in the crossover to the classical regime), a lot turns on 
these developments. 
 
In what follows, we first explain the decoherence problem, 
and describe some models used both for this
problem and for more general discussions of quantum relaxation.
We also describe the connection to 
experiments, particularly those in magnetic
and superconducting systems. In both cases we explain why the most
severe source of decoherence in the low-$T$ limit will often be the 
``spin bath'' 
environment of localised
modes (defects, spin impurities, nuclear spins) 
coupling to the collective coordinates (magnetisation, flux)
of interest. 

It is then but a small step to a discussion of the quantum 
measurement problem, for the simple reason that one can also use 
superconducting SQUIDs and nanomagnets as measuring systems themselves. Thus
we also grasp the nettle and consider models in which the entire set-up
(quantum system, apparatus, and environment) is quantised. This not only allows 
us to calculate the quantum dynamics of such a set-up, but also to discuss, in
a more quantitative way than usual, the measurement problem itself. These same
models also describe a pair of coupled mesoscopic qubits- and we discuss why decoherence typically has a much worse effect on the mutual coherence between 
these than it would on each qubit by itself.


{\bf Decoherence Puzzles}:
The basic idea of decoherence as a solution to the measurement problem has 
been around for some 
50 years \cite{decoOLD,omnes}; and as a physical phenomenon it has
been noted since before Quantum Mechanics. At first glance it seems 
almost trivial- most physicists and chemists are familiar with the 
way in which the phase interference in a ``2-slit'' kind of experiment
is destroyed.
So, why care about decoherence now? The short answer is that
we now need to know rather urgently what mechanisms cause decoherence, 
particularly at the mesoscopic and macroscopic scales, or in entangled systems
like quantum computers- and
we find that there are many basic things that we don't understand.
One way to introduce the problems is
in the form of a set of 
``decoherence puzzles'' or questions, which 
are clearly down-to earth and within the purview of physicists and
chemists. Here are a few (some of intense current interest):

{\it (1) Mechanisms of Decoherence }:
What are the mechanisms of decoherence in nature and how do they
relate to those governing quantum relaxation ? This problem is
very difficult because decoherence is extraordinarily sensitive
to the different environmental couplings in nature.
Thus {\em ad hoc} theoretical models, devised for their solvability
rather than their realism of generality, will not do here.
We are entering a particularly exciting time for
this puzzle, since experiments are now directly addressing coherence
phenomena at the mesoscopic and macroscopic levels.

{\it (2) Measurement Puzzle} : Does decoherence ``solve'' the
measurement problem (as many \cite{decoOLD,omnes,complex} 
have advocated)? How then  
must we
understand measurement operations in systems where macroscopic
coherence exists ?
 
{\it (3) Dynamics of Decoherence} : How does decoherence (in the relevant 
components of the relevant density matrix) evolve in time? 
How does decoherence
affect N-particle entanglement (eg, in an $N$-qubit quantum
computeri \cite{QC}) and how do N-particle decoherence rates depend on $N$ ?
How can the dynamics be controlled/tested experimentally ?

{\it (4) The $\tau_{\phi}$ puzzle}: Why, in defiance of very
basic theory \cite{tau_ph_th} does the decoherence rate $\tau_{\phi}^{-1}
(T)$ for electron propagation in mesoscopic conductors appear to
saturate \cite{tau_ph} to a finite constant at low $T$ ? Does a 
similar saturation exists in superconductors or magnets ? How
can $\tau_{\phi}^{-1} (T \rightarrow 0)$ be controlled ?

{\it (5) Quantum-Classical Crossover}: This is induced by raising
either temperature,
or external fields, or temperature.
The environment inducing the transition
is usually neither in internal equilibrium nor in equilibrium with the system
making the crossover. Yet this  departure from equilibrium 
is actually crucial (in, for
example, many discussions of quantum measurements); how can it be treated
properly?

{\it (6) High Energies}: What is the role of decoherence in
the early universe \cite{deco-early} 
(another possible quantum-classical crossover),
or in black hole physics \cite{deco-bh} ? 
Can high-energy sources of decoherence
up to or beyond the Plank scale have an effect at low energies 
\cite{deco-planck}? 
How can this be tested experimentally ? 

Note that these puzzles are all more or less related,
and that they span many different areas in physics (with
obvious repercussions in physical chemistry). It is also clear that the
first puzzle, about the mechanisms governing decoherence, is quite central,
and is the sort of bread-and butter question that physicists 
and chemists deal
with every day. Perhaps the first thing to understand is why it is possible
to approach this puzzle in a unified way (rather than by piecemeal discussion
of many different systems), and so we start with this.

\section{Spin Baths and Oscillator Baths}

Theorists working on the quantum dynamics of
open systems like to have models that are both realistic (realistic
enough to survive experimental tests on particular physical systems)
and also general (so that the results are generic, or at least 
apply to a large class of physical systems). One way of deriving such 
such models is to divide the quantum environment into extended and
localised modes, and then proceed as follows:


(i) {\bf Extended Modes}: 
In a box of volume $V$ containing $N_o \sim {\mathcal O}(V)$
extended modes below an ultraviolet cutoff $E_c$, the coupling
between these modes and some ``central system'' (collective
coordinate, etc) of interest is necessarily 
$\sim {\mathcal O} (N_o^{-1/2})$.
Then for large $N_o$ one can treat perturbatively and map to an 
environment of oscillators \cite{Feynman,cal83}. 
The general model is then described by
a Hamiltonian $H=H_o (P,Q)+H_{int}^{os}+H_{os}(\{ {\bf x}_q, {\bf p}_q \}) $ where $H_o(P,Q)$ is the
Hamiltonian of the ``central system'' we are interested in, having canonical 
coordinates $P,Q$, and the oscillator terms are
\cite{Feynman,ajl84}
\begin{equation}
H_{os} = \frac{1}{2} \sum_{q=1}^{N_o} m_{q}
( \dot{{\bf x}}_{q}^{2}
+ \omega_{q}^{2} {\bf x}_{q}^{2})
\label{hosc}
\end{equation}
\begin{equation}
H_{int}^{os} = \sum_{q=1}^{N_o}
[F_{q} (Q) {\bf x}_{q} + G_{q} (P) {\bf p}_q ]
\label{hint}
\end{equation}
where the $\{ {\bf x}_q,{\bf p}_q \}$ are canonical coordinates 
for the extended environmental modes (with 
$q=1,2, \ldots N_o$). The linear couplings $F_q, 
G_q \sim  {\mathcal O} (N_o^{-1/2})$,
so that the bath modes have individual effects
$\sim  {\mathcal O} (1/N_o)$ on the central system - summation over these
gives a well defined thermodynamic limit independent of $N_o$ as 
$N_o \rightarrow \infty$. 

The derivation of oscillator bath models generally uses some kind of 
Born/Op\-penheimer-like assumption that one can 
separate ``slow'' system modes 
$(P,Q)$ and ``fast'' environmental modes $\{ {\bf x}_q, {\bf p}_q \}$; such
derivations are of course very familiar to quantum chemists.  
To be valid they require that 
\begin{equation}
V_{qq'}(P,Q) \ll (\omega_q - \omega_{q'})
\label{BO}
\end{equation}
(or an appropriate generalisation to finite temperature $T$) for all 
$q$, $q'$, 
$P$, $Q$ of interest, where the matrix element at a given $P, Q$ is given, in 
terms of the full multi-dimensional coordinate space ${\bf X}$ of the environment, by
\begin{equation}
V_{qq'}(P,Q) = \int \, d{\bf X} \, \phi_q^*({\bf X}) \, \hat{H}_{int}^{os} 
\, \phi_q({\bf X}) 
\label{Vqq}
\end{equation} 
where the wave-functions $\phi_q({\bf X})$ are the eigenstates of the 
environmental Hamiltonian (for more details see, eg., Caldeira and Leggett
\cite{cal83}). Note that this restriction is not nearly as bad as it seems 
since one has considerable freedom in the choice of the environmental modes-
indeed the $\{ \omega_q \}$, as well as the couplings in (\ref{hint}), 
typically depend on 
$T$ (because, eg., one chooses a $T$-dependent effective Hamiltonian for the
environment).

However, there are obvious cases where the assumption (\ref{BO}) 
breaks down. 
In the case of extended environmental modes a well-known example is the Quantum
Hall system, where the environmental Landau levels are all degenerate- and the 
effect of $H_{int}^{os}$ is typically to break this degeneracy \cite{PS94}.
However the breakdown is much more common in the case of localised 
environmental modes, to which we now turn.


(ii) {\bf Localised Modes}: 
In the same volume $V$ there will also be be many localised
environmental modes, associated with defects, impurities, paramagnetic
spins and the nuclear spin bath. Each of these has a finite Hilbert space
in the energy range $ < E_c$, and can thus be mapped to a spin degree of
freedom, very often a spin$-1/2$, or two-level system. Some set 
$\{ \mbox{\boldmath $\sigma$}_k \} $ of these with $k=1,2, \ldots N_s$,
will also couple to the central system, giving extra terms
$H_{int}^{sp}+H_{sp}$ in the total Hamiltonian, where \cite{PS00}
\begin{equation}
H_{sp} = \sum_{k}^{N_s} {\bf h}_k \cdot \mbox{\boldmath $\sigma$}_k
+ \sum_{k,k'}^{N_s} V_{kk'}^{\alpha \beta} 
\sigma_{k}^{\alpha} \sigma_{k'}^{\beta}
\label{H_sp}
\end{equation}
\begin{equation}
H_{int}^{sp} = \sum_{k}^{N_s} \left[ F_{k}^{\|} (Q) \sigma_{k'}^{z}\;+\;
(F_{k}^{\bot} (P,Q) \sigma_{k'}^{+} + h. c.) \right] 
\end{equation}
The important point here is that there is no requirement that the
couplings $\{ F_{k}^{\|}, F_{k}^{\bot} \}$ be ${\mathcal O}
(N_s^{-1/2})$;
indeed very often they are independent of $N_s$ ! A simple 
example is the hyperfine coupling, crucial in magnetic systems since
it couples the macroscopic magnetisation to the nuclear spins- this is 
obviously independent of $N_s$. 
Note that the couplings are not necessarily small either; for example, 
hyperfine
couplings can be as large as $0.5 K$ each, and individual exchange
couplings to paramagnetic impurities can be much larger again. Now 2 
crucial points arise, viz.,

(i) The {\em intrinsic} dynamics of the environmental modes, 
coming from $H_{sp}$ in (\ref{H_sp}) is now 
governed by energy scales
$\vert {\bf h}_k \vert$, $\vert V_{kk'}^{\alpha \beta} \vert$ which are often
much weaker than the couplings $ \{ F_{k}^{\|}, F_{k}^{\bot} \}$, putting us 
then in an ``anti- Born Oppenheimer'' limit, where the 
environmental modes are strongly slaved to the system coordinates $P,Q$. This
is because the modes, having localised wave-functions, usually couple very 
weakly to each other (via, eg., residual dipolar couplings). We may again take  magnetic systems as an example, where the internuclear coupling 
$\vert V_{kk'}^{\alpha \beta} \vert \sim 10^{-8}~K$ typically, ie., up to
8 orders of magnitude smaller than the hyperfine couplings
$ \{ F_{k}^{\|}, F_{k}^{\bot} \}$. 

(ii) The Hilbert space of an oscillator is infinite-dimensional and thus 
quite different from that of a spin (or any environmental mode with only a few 
discrete levels). This would not be important if the 
weak-coupling assumption (\ref{BO}) were satisfied; but in the common case
where it is not, 
one has to use spin propagators in dealing with the environmental
modes. 


Readers working in different fields will recognise many of their favourite
models in these general forms (quite how general they are 
has been discussed at some length in the
literature \cite{ajl84,PS00,ajl87,weiss}). The
models are completely specified once (a) the Central system Hamiltonian has 
been given, and (b) the various couplings (or more usually, their 
distributions over $q$ and $k$) are known. At this point two  
simplifying features usually emerge, viz.; 

(a) The distribution of 
couplings to the environmental modes shows a fairly typical pattern,
(see Fig. \ref{fig1}). Coupling to oscillator systems is usually much  
weaker at low energies and temperatures, as the available phase
space disappears, except for the case of ``Ohmic baths''
such as conducting electrons in a metal (and even this coupling rapidly 
disappears 
at low temperatures if the metal goes superconducting, so that the electronic
spectrum becomes gapped). Phonons and photons 
in particular have very weak low frequency couplings to most central 
systems of interest. On the other hand, the coupling to spin baths is often 
much {\it stronger} at low energies -- this is particularly true for 
coupling to 
nuclear spin systems, which becomes much stronger as one goes to energies 
$\sim E_o = [\sum_k (F_k^{\parallel})^2 ]^{1/2}$. For this reason, 
spin baths are often much more serious sources of low-$T$
decoherence than oscillator baths.

(b) At low temperature the Hamiltonian of
the central system is often almost trivial. In microscopic systems like
2-level atoms this may not come as a surprise (although an atom is of course 
already a very complex many-body system). In mesoscopic or macroscopic 
quantum systems the central Hamiltonian is typically describing a 
collective coordinate which may have a quite large ``mass'' or 
``inertia'', 
but whose dynamics 
simplifies at low energy because it is restricted by some confining potential.
Thus one ends up with a few ``canonical models'' \cite{toms} which have
been applied to many different physical situations- these include a 
1-dimensional harmoic oscillator \cite{grabert}, a simple tunneling 
potential \cite{cal83,weiss}, and a variety of problems in which the central 
Hamiltonian $H_o$ describes one or more 2-level systems. In fact the problem 
of a single ``central spin'' (or ``qubit'') is so important that we pause to 
sketch a few details.


(iii) {\bf Spin-Boson \& Central Spin Models:}  If we couple a central
two-level system to either an oscillator or a spin environment, 
we get two very useful
models.
In the case of an oscillator bath we get the spin-boson Hamiltonian
\cite{ajl87,weiss}
\begin{equation}
H_{SB} = \Delta \hat{\tau}_x + \xi \hat{\tau}_z + 
\sum_q [c_{q}^{\|} \hat{\tau}_z + (c_{q}^{\bot} \hat{\tau}^{+} + h.c)]
+H_{os}
\end{equation}
with $H_{os}$ given in Eq. (\ref{hosc}). The spin 
$ \mbox{\boldmath $\tau$} $ is in a longitudinal bias field
$\xi$, with tunneling matrix element $\Delta$ between states
$| \uparrow \rangle$ and $| \downarrow \rangle$; the couplings
$c_{q}^{\parallel}, c_{q}^{\perp} \sim {\mathcal O} (N_o^{-1/2})$. This model
has been widely used \cite{ajl87,weiss,PS96} 
to discuss systems like SQUIDs, or nanomagnets,
or other 2-state mesoscopic systems coupled to delocalised
excitations
like electrons, phonons, magnons, photons, etc (it was originally 
introduced to discuss problems like nucleons \cite{wilson} or the 
Kondo problem \cite{kondo}, involving microscopic 2-level systems). 

If we couple to a spin bath we get the central spin Hamiltonian
\begin{equation}
H_{CS} = \left( \Delta \hat{\tau}_+ 
\exp^{i \sum_k  \mbox{\boldmath $\alpha$}_k 
\cdot \mbox{\boldmath $\sigma$}_k} \;+\;
H. c. \right) 
+ \sum_k \omega_{k}^{\|} \hat{\tau}_z \sigma_{k}^{z} + H_{sp}
\label{H_CS}
\end{equation}
with $H_{sp}$ given in Eq. (\ref{H_sp}). Here, as noted above,
$\omega_{k}^{\|}$ and $\mbox{\boldmath $\alpha$}_k$ 
are often independent of $N_s$. This model has
been used to describe systems like nanomagnets \cite{PS96} or 
SQUIDs \cite{PS00,mooij,sq00} coupled 
to nuclear and paramagnetic spins. 

The non-diagonal couplings $c_{q}^{\bot}$ and 
$ \mbox{\boldmath $\alpha$}_k$ in these models are typically smaller than
the diagonal couplings - for a complete discussion of this point, see
Ref. \cite{PS00}. 
However, one should be careful not to drop them when discussing
decoherence (see below).

These 2 models can be combined (ie., a central spin couples simultaneously
to spin and oscillator baths- this often happens 
in reality \cite{PS00,PS96}),
and they can also be generalised to discuss 2 or more spins coupled both 
amongst each other and to an environment -- we will use such models in sections
5 and 6. 
 

(iv) {\bf Averaging over the environment: } To complete the discussion of these 
models for open quantum systems one must 
specify how to average over or ``integrate out'' the environmental modes. 
This
has been extensively reviewed \cite{weiss,PS00,FeynH} so we 
simply recall the 
results here. For both spin and oscillator baths the best way 
is to use path integral methods (lending themselves particularly to 
strong-coupling and tunneling problems). For the oscillator bath a common 
method is to write the proagator for the density matrix as a functional 
integral over the ``influence functional'' 
${\mathcal F}[q,q']$, in the form
\cite{weiss,FeynH}: 
\begin{eqnarray}
& & K(Q_f,Q_f^{\prime};Q_i,Q_i^{\prime};t) = \nonumber \\
& & \int^{q(t) = Q_f}_{q(0) = Q_i}
{\mathcal D}q(\tau)
\int^{q^{\prime}(t) = Q^{\prime}_f}_{q^{\prime}(0) = Q^{\prime}_i}
{\mathcal D}q^{\prime}(\tau)
e^{i(S_o[q] - S_o[q^{\prime}])} {\mathcal F}[q(\tau),q^{\prime}(\tau)]
\label{F[qq]}
\end{eqnarray}
where $S_o[q]$ is the action of the free system (no environment) and 
${\mathcal F}[q,q^{\prime}]$ is a functional over all possible paths
$q(\tau)$, $q^{\prime}(\tau)$ of the system. 
${\mathcal F}[q,q^{\prime}]$ is usually a rather complex
object, and the functional integration over it is not easy either- what makes 
it possible at all is the weak coupling to the oscillators. The only cases 
for which the dynamics of $\hat{\rho}(t)$ have been established so far are 
when the central system is either an oscillator itself, a central spin (the 
spin-boson model) a pair of spins (the PISCES model discussed below), or a 
particle moving either in zero potential or hopping on a lattice, in the 
presence of oscillators. 

For the spin bath a nice simplifying feature appears, originating from the 
finite Hilbert space associated with each spin bath mode. One begins by
quantizing the bath spins along a $\hat{z}$ axis, and classifying all the 
$2^N$ bath states according to their total polarisation $M$ along this 
axis.
Instead of 
functional integrals over all possible paths of the system, one has just
ordinary integrals over the density matrix $\hat{\rho}^{(o)}(t)$ of the free 
system \cite{PS00}:
\begin{equation}
\rho(Q,Q^{\prime};t) = \hat{{\mathcal A}}^T(\phi) \hat{{\mathcal A}}^O(y)
\hat{{\mathcal A}}^B(\epsilon,M) \; 
\rho^{(o)}(\phi,y,\epsilon,M; Q,Q^{\prime};t)
\label{rho(t)}
\end{equation}
where the parameters in the free density matrix $\rho^{(o)}(t)$ now become 
simple functions of (i) a phase variable $\phi$ associated with rotations of 
bath spins induced by transitions of the central system (ii) a similar 
variable $y$ associated with spin bath precession in between central system 
transitions, and (iii) the energy bias $\epsilon$ acting on the bath spins,
and the polarisation $M$ of the bath. The averages 
$ \hat{{\mathcal A}}^T(\phi)$, $\hat{{\mathcal A}}^O(y)$,
and $\hat{{\mathcal A}}^B(\epsilon,M)$ are just simple weighted 
integrations over 
these variables (with weighting functions depending on the couplings to the
bath and the bath temperature). The replacement of functional integrals
by ordinary integrals makes the calculation of dynamics for the spin bath
much simpler than for oscillator baths- in fact one can usually find analytic 
answers. One sees this clearly  
in the example of the central spin model (\ref{H_CS}) (see ref \cite{PS00}
and below).

\section{Decoherence for single ``qubits'': Experimental Implications}

Perhaps the best way to understand a formalism is to look at some important
application of it to a real system. We therefore begin with a very important
one, to experiments on mesoscopic SQUIDs and magnets- both systems are strong 
candidates for qubits in some future quantum computer. 
In fact many such experiments have 
been done in the last few years, which we can somewhat arbitrarily divide up as 
follows: 

(i) Experiments 
claiming observation of coherence, or at least interference, between
mesoscopic quantum states, either in SQUIDs \cite{fried,Wal} 
or quantum magnetic systems \cite{awsch,barco,luis}. 

(ii) Experiments on incoherent tunneling at the mesoscopic or macroscopic 
scales, also in superconductors 
\cite{webb,clarke,lukens} and magnets 
\cite{barb,wern,giord,tunn96,ohm98}), done in 
response to theory \cite{cal83,ajl87,amb,garg91,IJMPB,sta92,PS93} which 
indicated that even in the presence of a vast number of environmental modes or 
``microstates'', such tunneling was possible. 

In fact the incoherent tunneling experiments were very successful- and 
amongst other things 
they indicate that the theoretical treatment of the environment works pretty 
well. Thus, the SQUID tunneling experiments of Clarke et al. \cite{clarke}
gave amazingly good agreement with the Caldeira-Leggett \cite{cal83} 
predictions for
electronic ``oscillator bath'' dissipation in SQUIDs. Recent experiments on 
crystalline ensembles of molecular magnets \cite{ohm98,wer00} (including 
direct verification of the role of nuclear spins, by isotopic variation
\cite{wer00}) also agreed quantitatively 
with the theoretical predictions \cite{PS96,PS93,PS98} for 
nuclear spin-mediated incoherent tunneling in these systems. Thus at least for 
incoherent tunneling, the theory seems to work. 

However, as emphasized long ago in the measurement theory literature,
interference between 2 ``classical'' (ie., mesoscopically or macroscopically different) states is a much more sensitive test of our understanding of
decoherence mechanisms. In the same way it severely tests
our understanding of superconductors and magnets, and indeed of many-body
physics as a whole. The coherence experiments in SQUIDs and magnets 
are usually analysed
using the spin-boson and/or central spin models just given, because in both 
cases the collective coordinate dynamics is governed by a 
``2-well'' potential
at low energies when $k_BT < \omega_o/2 \pi$, where $\omega_o$ is 
roughly the same as the gap from the 2 lowest states in this potential, to 
the higher levels. In mesoscopic magnets this ``spin gap'' 
may easily be $10~K$; 
in SQUIDs it is the Josephson plasma frequency, which can be made as large as 
$1-2~K$ (incidentally, $\Omega_o$ does {\it not} 
decrease exponentially with the size of
the system, as has sometimes been argued \cite{omnes}; in magnets it may hardly
depend on system size at all).

In this section we discuss the mechanics of decoherence for problems like
these, returning at the end of the section to make some brief comments about
the experiments themselves.
Note what is meant by coherence in the case of
these 2-level systems -- if, say, we start with the 2-level system in the pure
state $\vert \uparrow \rangle$, then a purely coherent time evolution 
thereafter is described by the ``free'' 
density matrix $\rho_{ij}^{(o)}(t)$, 
where (see Fig. \ref{fig2}):

\begin{equation}
\rho^{(o)} (t) = \left( 
\begin{tabular}{cc}
$  \frac{1}{2} \left( 1 + \cos \left( 2 \Delta t \right) \right) $  &
$ \frac{i}{2} \sin \left( 2 \Delta t \right) $ \\ \\ 
$ - \frac{i}{2} \sin \left( 2 \Delta t \right) $ &
$  \frac{1}{2} \left( 1 - \cos \left( 2 \Delta t \right) \right) $
\end{tabular}
\right)
\label{RHOo}
\end{equation} 

The job of the theory is to predict the evolution of $\rho_{ij}(t)$ in the 
presence of the environment -- coherence between the states 
$\vert \uparrow \rangle$ and $\vert \downarrow 
\rangle$ is then described by the
{\it off-diagonal} matrix elements $\rho_{12}$ and $\rho_{21}$. 
So far the experiments have not directly tested any decoherence calculations,
and indeed the magnetic coherence experiments are controversial- see below. 
However they soon will, and so in this section we discuss what theory says 
for the decoherence dynamics of these off-diagonal matrix elements.


(i) {\bf Spin-Boson Model: }  We include only the diagonal coupling
$c_q^{\parallel}$ for simplicity, and assume to be specific that the coupling 
is Ohmic and weak- formally this means that the
Caldeira-Leggett \cite{cal83} spectral function $J(\omega)= \pi \alpha \omega$
with $\alpha \ll 1$. 
This model then applies directly to the electronic 
contribution to decoherence in SQUID tunneling between flux states
$\pm \phi_m$ with charging energy $E_c$ and a given Q-factor 
\cite{amb}; where $\alpha = 
( 16 \phi_m^2 \omega_0 /E_c ) Q^{-1}$. 
The dynamics for this model is fairly well-known 
\cite{ajl87,weiss}; in the weak-coupling limit one has standard exponental 
damping of the oscillations, with diagonal and off-diagonal decay rates
given respectively by \cite{ajl87,Gri97,PK90}: 
\begin{equation}
\Gamma_{11}(\Delta_r, \xi) = (\Delta_r/E)^2 \gamma(E)
\label{g11}
\end{equation}
\begin{equation}
\Gamma_{12}(\Delta_r,\xi) = 
{1 \over 2E^2}[2\xi^2 \gamma(0) + \Delta^2 \gamma(E)]
\label{g12}
\end{equation}
where the rate function $\gamma(E)$ is
\begin{equation}
\gamma(E) = \pi \alpha E \coth(E/k_BT)
\label{gE}
\end{equation}
and $E = (\xi^2 + \Delta_r^2)^{1/2}$ with a renormalised tunneling
matrix element $\Delta_r$.
A somewhat peculiar point is that the {\it size} of the off-diagonal matrix 
elements is also reduced, by a factor $\Delta_r / \Delta \sim 
(\Delta/E_c)^{\alpha/1-\alpha}$, which explicitly contains the high-energy 
cut-off in the oscillator bath spectrum (Fig. \ref{fig3}); similar
effects can also be seen for a free particle \cite{Hak85}.

We can think of these results as calculations of the rates $T_1^{-1}
\equiv \Gamma_{11}$ and $T_2^{-1} \equiv \Gamma_{12}$, with $T_2 = 2T_1$
for the symmetric (zero bias) case $\xi =0$. Thus for a spin weakly coupled
to an oscillator bath, the decoherence rate $T_2^{-1}$ is simply related
to the longitudinal rate $T_1^{-1}$. This is partly a reflection of the 
simplicity of the decoherence mechanism here- the longitudinal coupling
$c_q^{\parallel}$ couples $\tau_z$ to a ``fluctuating bias field'' (the 
oscillators).

Anyone familiar with the usual arguments for the destruction of off-diagonal 
matrix elements as a solution to the ``measurement problem'' ought to be 
suspicious of this simple 
relationship between relaxation and decoherence in the 
spin-boson model. In fact it is too simple, indeed 
rather exceptional- as a general rule one 
expects no particular connection between them. Results from the
spin-boson model can thus be misleading, so we 
now turn to the central spin results. 


(ii) {\bf Central Spin model}: From Fig. 1 we learn that most candidates 
for a ``qubit'' will begin to feel the deleterious effects of the spin bath 
at sufficiently low $T$. By itself the coupling
$\omega_k^{\parallel}$ causes no decoherence at all- except for a rather weak
contribution coming from the internal dynamics of the spin bath
(driven by $V_{kk'}$). The 2 main sources of decoherence are (i) the precession
of the spins in the spin bath between transitions of the central spin, 
driven by the term $\omega_k^{\perp}$, and (ii) transitions in the spin bath
caused by the flipping of the central spin (coming from the 
$\mbox{\boldmath $\alpha$}_k$ term- the probability for 
$\mbox{\boldmath $\sigma$}_k$ to flip during a single central spin transition 
is $\vert \mbox{\boldmath $\alpha$}_k \vert^2$ when 
$\vert \mbox{\boldmath $\alpha$}_k \vert \ll 1$. Both of these decoherence 
mechanisms \cite{PS93,PS00} involve
the accumulation of a random phase by the environment, which when integrated 
over to obtain the reduced density matrix of the central spin, lead to 
decoherence. The extent of this decoherence is quantified using the 
parameters
\begin{equation}
\lambda = {1 \over 2} \sum_k \vert \mbox{\boldmath $\alpha$}_k \vert^2
\label{lambda}
\end{equation}
\begin{equation}
\kappa = {1 \over 2} \sum_k (\omega_k^{\perp}/\omega_k^{\parallel})^2
\label{kappa}
\end{equation}
(we assume $\vert \mbox{\boldmath $\alpha$}_k \vert, 
\vert \omega_k^{\perp}/\omega_k^{\parallel}\vert \ll 1$). Roughly speaking, 
these dimensionless numbers tell us the mean phase accumulated by each of 
these mechanisms whilst it operates. Thus we expect strong decoherence if 
$\lambda >1$ and/or $\kappa >1$. 
To substantiate this we give the results for the 
off-diagonal matrix element $\rho_{12}(t)$ 
of the central spin density matrix in each case
(for the diagonal matrix elements see refs. \cite{PS93,PS00}),
since these matrix elements quantify decoherence effects:


(a) {\it Topological Decoherence limit}: In this limit the main source
of decoherence is the transitions in the spin bath caused by a transition in
the central system (``co-flip processes''); 
the time-dependent perturbation on 
the bath spins, coming from the central spin transition, changes the spin 
bath state, and each such transition gives a random phase (mean value 
$\lambda$ ) to the bath. One then has 
\begin{equation} 
\rho_{12}(t) = \hat{{\mathcal A}}^T(\phi) 
\rho_{12}^{(o)}(\phi;t)
= \sum_{\nu = -\infty}^{\infty} \int_0^{2\pi} e^{i\nu \phi -4\lambda\nu^2}
\rho_{12}^{(o)}(\phi;t)
\label{rho1}
\end{equation}
which is easily evaluated using $\rho_{12}^{(o)}(\phi;t) = (i/2)
\sin(\Delta(\phi)t)$,
where the phase-dependent tunneling matrix element is found \cite{PS93} to be
$\Delta(\phi) = \Delta \cos(\phi)$. Only
terms with odd winding number $\nu$ contribute, and one finds that
\begin{equation}
\rho_{(12)}(t) = \sum_{\nu} e^{-4\lambda(2\nu +1)^2} J_{2\nu + 1}(\Delta t)
\label{rho2}
\end{equation}
In the case $\lambda \rightarrow 0$ (ie., the bath decouples from the 
central spin) this just 
goes back to the free spin result in (\ref{RHOo}). If $\lambda \sim O(1)$
or greater, we have $\vert \rho_{12}^{(o)}(t) \vert \sim O(e^{-\lambda})$,
ie., exponential suppression of the off-diagonal matrix element 
(cf. Fig. \ref{fig4})!


(b) {\it Orthogonality Blocking limit}: In this limit ``co-flip'' processes
are unimportant, but the bath spin precession between each central spin 
transition now adds a random phase to the bath part of the total 
wave-function- when integrated over this also leads to decoherence, in much 
the same way as with topological decoherence. This mechanism becomes
particularly important when the couplings $\{ \omega_k^{\parallel} \}$ are
large compared to $\Delta$, in which case it is important to distinguish the 
different polarisation groups in the spin bath. We then have 
\begin{equation}
\rho_{12}^{(o)} (t) = \hat{{\mathcal A}}^O(y) 
\rho^{(o)}_{12} (y;t) 
\rightarrow  \sum_{M= -N}^{N} w_M(T) \langle \rho_{12}^{(o)}
(t,\Delta_M(y))
\rangle
\end{equation}
where the weighting function $w_M(T)$ tells us the fraction of spin baths, in
a thermally weighted ensemble, to be found in the $M$-th polarisation
group, the phase-weighted matrix element is now $\Delta_M(y) = \Delta 
J_M(\sqrt{2\kappa y})$ and the average is
\begin{equation}
\langle \rho_{12}^{(o)}(t,\Delta_M(y)) \rangle = 
\delta_{M,0} \int_0^{\infty} dy e^{-y} \rho_{12}^{(o)}(t,\Delta_M(y))
\label{orth}
\end{equation}
This result is rather remarkable- since the fraction of spin baths in an 
ensemble which find themselves in polarisation group $M=0$ is 
$\sim (2\pi N)^{-1/2}$, we have only a tiny contribution to the off-diagonal 
matrix elements! The physical reason for this result is actually very simple-
for coherence to proceed, the state of the nuclear bath must not change between 
initial and final states. But it is easy to see that in the present case
this can only happen if $M=0$, because otherwise the system cannot make 
resonant transitions (recall that  $\{ \omega_k^{\parallel} \} > \Delta$).
Consider, eg., the case $M=1$. If no bath spins flip during the transition, then
afterwards the central spin sees $M= -1$, ie., the total coupling energy,
acting here like a longitudinal bias,
has changed by roughly $2 \omega_k^{\parallel}$ (for more detail on this 
physics see ref. \cite{PS96}, Appendix B).


(iii) {\bf Experimental Implications}; Amongst other things these results 
demonstrate the well-known fact that there is no 
necessary connection between the decay of diagonal matrix elements
and those of off-diagonal ones. This may seem obvious to those familiar 
with either measurement theory or NMR. In NMR one often sees a 
huge difference in behaviour between 
$T_1$ and $T_2$; but coherent dynamics exists only 
over timescales $< T_2$ (off-diagonal
matrix elements), and {\it not} for 
$T_2 < t<T_1$ (even though oscillations in the
diagonal density matrix elements may still be strong). 

However as far as we are aware this point 
has not been noted up until now in the context of 
macroscopic coherence experiments. In fact all of these experiments  
have either looked at densities of states using thermodynamic measurements
\cite{luis},
or else at the absorption spectrum of the relevant system (SQUID or magnet)
as a function of frequency and applied field 
\cite{mooij,fried,awsch,barco}.
We would like to emphasize that none of these experiments is then directly 
demonstrating coherence- one can easily envisage a situation in which 
resonant structure in either the thermodynamic density of states or the 
resonant absorption is coming entirely from the diagonal matrix elements.
Certainly the SQUID experiments of van der Wal et al. \cite{mooij}, which look 
at ``ground state tunneling'' (unlike the excited state tunneling experiments 
of Friedman et al. \cite{fried}), 
give strong evidence 
that such coherence may soon be demonstrated directly- but they do not
in themselves constitute such a demonstration. As for the magnetic experiments 
so far, two of these experiments (both highly controversial 
\cite{awschC,barcoC}) have been done on large ensembles of disordered 
nanoparticles, and although 
the specific heat experiment \cite{luis} has been done on an ordered set of 
nanomolecules, it does not probe the magnetic dynamics directly.

However what is certainly true is that there is nothing in the decoherence
results described so far that rules out macroscopic coherence. Decoherence 
coming from the oscillator bath can easily be made very small for either a SQUID
(where one can make $\alpha \ll 1$) or an insulating magnet (where phonons give 
very little low-energy decoherence). The results for spin bath decoherence
seem more worrying but only if $\lambda$ and/or $\kappa$ are large. In a SQUID
the best strategy to avoid spin bath decoherence is to make $\Delta$ large
(as experimentalists are doing), control substrate decoherence, and possibly 
also to {\it increase} the size of the SQUID ring. In a magnetic 
system the best 
way to reduce decoherence effects turns out to be 
using strong transverse fields,
which not only increases $\Delta$ but also radically reduces $\kappa$; 
again, the theory indicates that coherence should be visible. We 
do not as yet have any reason to disbelieve this theory- in fact it has done 
rather well in predicting the tunneling characteristics of these large systems.
Thus the short-term prognosis looks good- it would seem reasonable to bet that 
direct evidence for single mesoscopic solid-state qubit operation will be found
in the next year or two. What then?

\section{Coupled Qubits, Quantum Measurements, and all that} 

The next step up from single qubits is to couple pairs of them and look at 
their ``mutual coherence'' or ``entanglement'', and how it is destroyed by 
decoherence processes. Ine the same way that one can write down general 
models for single qubits coupled to an environment (the spin-boson and central 
spin models) one can do the same for a pair of them- this is the 
``PISCES model'',
to which we come below. However it is interesting to realise that the same 
model also describes a measurement system, itself operating in the quantum
regime, and set up to probe a single qubit. This equivalence is not reflected
in the literature, where 2 quite different approaches coexist (somewhat 
uneasily!). 


(i) {\bf Quantum Measurements}: The quantum  
measurement puzzle is perhaps the most controversial of all those
associated with Quantum Mechanics. The first problem is of course to say what 
it is- this depends partly on one's view of ``physical reality'', as well as 
on how one formulates the theory. One simple approach begins by noting that 
that in an ``ideal'' measurement, the coupled 
``system-apparatus'' is supposed to evolve according to 
\begin{equation}
\Psi_{in} = \Phi_o \sum_{j} c_j \phi_j  \;\;\; \rightarrow \;\;\; 
\Psi_f = \sum_j c_j \Phi_j \phi_j
\label{mmt}
\end{equation}
leaving a ``post-measurement'' state $\Phi_j$ of the apparatus in 
$1 \leftrightarrow 1$ 
correlation with an initial state $\phi_j$ of the measured system (obviously 
we can also have $\Psi_f = \sum_j c_j \chi_j \Phi_j$, with $\chi_j = 
\sum_{ji}\alpha_{ji} \phi_i$). The puzzle is then simply that for the 
apparatus to function as intended, the states $\{ \Phi_j \}$ must be 
macroscopically different from each other- meaning that (\ref{mmt}) 
describes a superposition of macroscopically different states.

Answers to this puzzle vary widely. Many formulations of quantum mechanics
(including the Copenhagen interpretation) make measurements a primitive and 
unanalysable part of the theory, and insist that a measuring apparatus must
by definition be classical. The superposition $\Psi_f$ is then replaced by 
a mixture, the two being distinguished formally by their density matrices:
\begin{equation}
\rho_{ij}^{pure} = c_i c_j^*   \;\;\;\;\;\;\;\;\;\; (superposition)
\label{pure}
\end{equation}
\begin{equation}
\rho_{ij}^{mixt} = \vert c_i \vert^2 \delta_{ij} \;\;\;\;\;\;\;\;\;(mixture)
\label{mixt}
\end{equation}
Various arguments have been given that, FAPP (``For All Practical 
Purposes'' \cite{FAPP}), or even in principle, this manoeuvre is justified.  
Most of these start from the idea that the apparatus is extremely
``complex'', and so has many microstates, in a very large Hilbert space, 
associated with each of the ``macrostates'' describing 
a given outcome of a measurement \cite{omnes,complex}. 
It is also often argued that 
large amplification and very strong irreversibility (and/or chaotic dynamics) 
are associated with the time evolution of the apparatus during measurement. 
Hence, it is argued, no conceivable measurement performed on ``system $+$ 
apparatus'' could distinguish $\rho^{pure}_{ij}$ from $\rho_{ij}^{mixt}$
(it would require an operator simultaneously off-diagonal in all the relevant 
states!). Many of the early ``decoherence'' arguments concerning quantum 
measurements, going back some 50 years, make a similar point \cite{decoOLD}.
From now on we will refer to this set of arguments as the 
``orthodox argument''
for the denial of macroscopic quantum superpositions.

At first glance it appears to be a very powerful argument -- after all, the 
macroscopic world does appear to be classical, and the number of 
microstates to which a macroscopic ``pointer state'' couples is very large 
indeed. 
However the orthodox argument also suffers from 
a remarkable vagueness- no attempt 
is made to substantiate it by any detailed quantitative
theory, susceptible to experimental test on a real measuring system. 
One is reminded of the 1960's conceptual art movement 
(in which the point was not 
to create a work of art but rather to talk about what it would be like if 
completed). The problem  
with the argument is obvious- insofar
as it denies the existence and/or observability of mesoscopic
or macroscopic superpositions, it is in imminent danger of 
contradicting experiment (see previous
section) and certainly conflicts with a lot of well-established theory! 
But what precisely is the flaw in the orthodox argument?


(ii) {\bf Coupled qubits}: In the Quantum Computation 
literature \cite{QC}, on the other 
hand, the above orthodox argument is usually ignored completely (except for 
the occasional reference to decoherence as ``an important problem to be 
resolved'', occasionally accompanied by some rough calculation including a 
phenomenological ``noise term''). It is very hard to understand this attitude,
which entirely ignores the microstates so central to both the orthodox argument 
and to decoherence- after all, the whole point of 
quantum computation is to set up extremely 
delicate superpositions involving many qubits and a huge number of states. 
So how sensitive is the mutual coherence between qubits to the environment?  

Theoretical analysis of this question is simplified by the fact that quantum
computation algorithms can be set up by coupling pairs of qubits in various 
sequences \cite{QC}. This allows us to discuss decoherence in the measurement
process using the same model as we use for decoherence in the 
simplest quantum computations, a model involving 2 quantum systems plus the 
quantum environment. Let us now 
recall this model. 


(iii) {\bf Effective Hamiltonian}:
To address the above 
questions we consider a SQUID magnetometer or a nanomagnetic 
field sensor, in the quantum regime ($kT < \Omega_o/2 \pi$), set up so its 
2 low-energy  ``apparatus pointer macrostates'' $\vert \uparrow \rangle$, 
$\vert \downarrow \rangle$ couple diagonally to another 2-state system 
(see Fig. \ref{fig5}). 
Here we will assume the the measured system is also a 
mesoscopic or macroscopic SQUID or nanomagnet; the apparatus
pointer states will be described by a spin-$1/2$ variable $\vec{\tau}_1$, and 
the system states by a spin $\vec{\tau}_2$. 
Obviously such a set-up can also be used as a coupled qubit system if the 
parameters are adjusted suitably, and decoherence is sufficiently small in 
both of them.

We may represent 
the low-energy dynamics of this coupled system by a Hamiltonian which 
generalises to a pair of spins the spin-boson and central spin Hamiltonians 
given above. For simplicity we will consider here only delocalised 
environmental excitations (ie., ``microstates''); it is easy to generalise the 
following discussion, 
when necessary, to incorporate the coupling to a spin 
bath. Then we get the ``PISCES'' Hamiltonian \cite{PIS}
$H_{PISCES} = H_o^{P} (\mbox{\boldmath $\tau$}_1, \mbox{\boldmath $\tau$}_2)
+H_{int}+H_{os}$, with $H_{os}$ given by (\ref{hosc}), and 
\begin{equation}
H_o^P =
(\Delta_1 \hat{\tau}_{1}^{x} + \Delta_2 \hat{\tau}_{2}^{x} ) +
(\xi_1  \hat{\tau}_{1}^{z} + \xi_2  \hat{\tau}_{2}^{z} ) +
{\mathcal K}  \hat{\tau}_{1}^{z} \hat{\tau}_{2}^{z}
\end{equation}
\begin{equation}
H_{int} = \sum_q \left[ c_{q}^{(1)} \hat{\tau}_{1}^{z} + 
c_{q}^{(2)} \hat{\tau}_{2}^{z} \right] x_q
\end{equation}
To be specific in the following discussion we have made some 
restrictive assumptions, as follows:

(i) we assume only longitudinal couplings to the
oscillators. A more general model would have a coupling 
${\mathcal K}_{\alpha \beta} \, \hat{\tau}_{2}^{\alpha} \, 
\hat{\tau}_{2}^{\beta}$
between the 2 spins $\mbox{\boldmath $\tau$}_1$ and 
$\mbox{\boldmath $\tau$}_2$, but if the systems are mesoscopic, the
longitudinal couplings usually dominate the low-energy limit \cite{PIS}.
We assume only an oscillator bath- and to be even more specific we will 
assume a $T$- independent Ohmic coupling to the 2 spins.

(ii) We ignore time-dependence in the parameters- this is because we are 
interested here in environmental decoherence (as opposed to that caused by
this time-dependence, which is of course crucial to the actual operation of
a pair of qubits).  

A microscopic derivation of this Hamiltonian for 
the present situation is basically just the derivation of the effective 
couplings, and this we know how to do. For a single isolated SQUID 
one has a coupling to electronic excitations \cite{amb}, as well as less 
important couplings to phonons and photons-
and one has calculable couplings 
to a spin bath of nuclear and paramagnetic spins, as well as charge defects 
\cite{sq00}. In a similar way one can treat the 
coupling of a single quantum nanomagnet to electrons, phonons, and nuclear 
spins
in this low-energy limit \cite{PS96,tup97}. When two such systems are coupled 
one has to do a little more work -- the couplings to the various baths cause
a renormalisation of the effective coupling between the two spins -- 
but this can also be done for these systems \cite{PIS}. 

\begin{figure}[h]
\epsfxsize=12.0cm \epsfysize12.0cm
\epsfbox{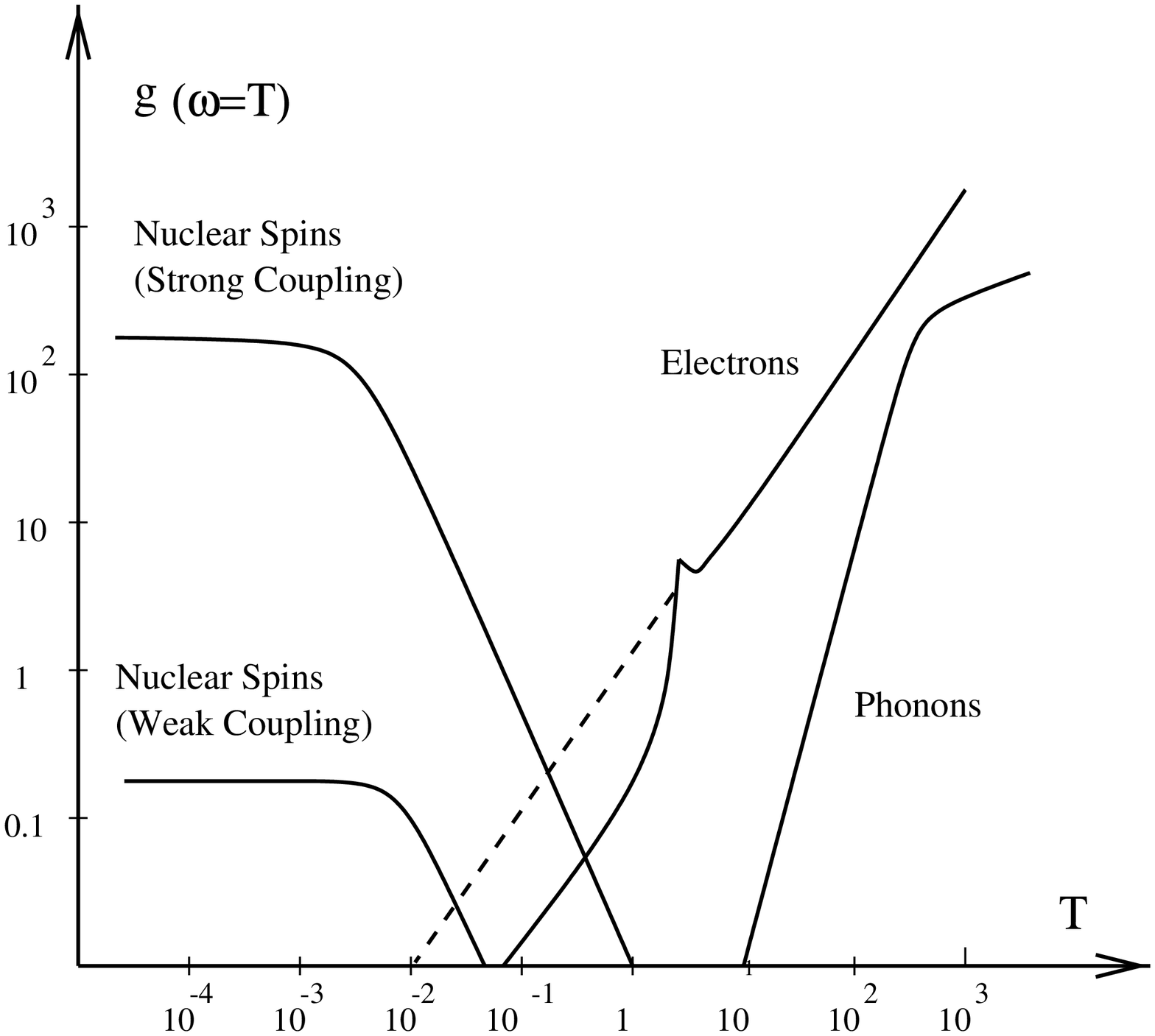}
\caption{
}
\label{fig1}
\end{figure}

\begin{figure}
\epsfxsize=12.0cm \epsfysize12.0cm
\epsfbox{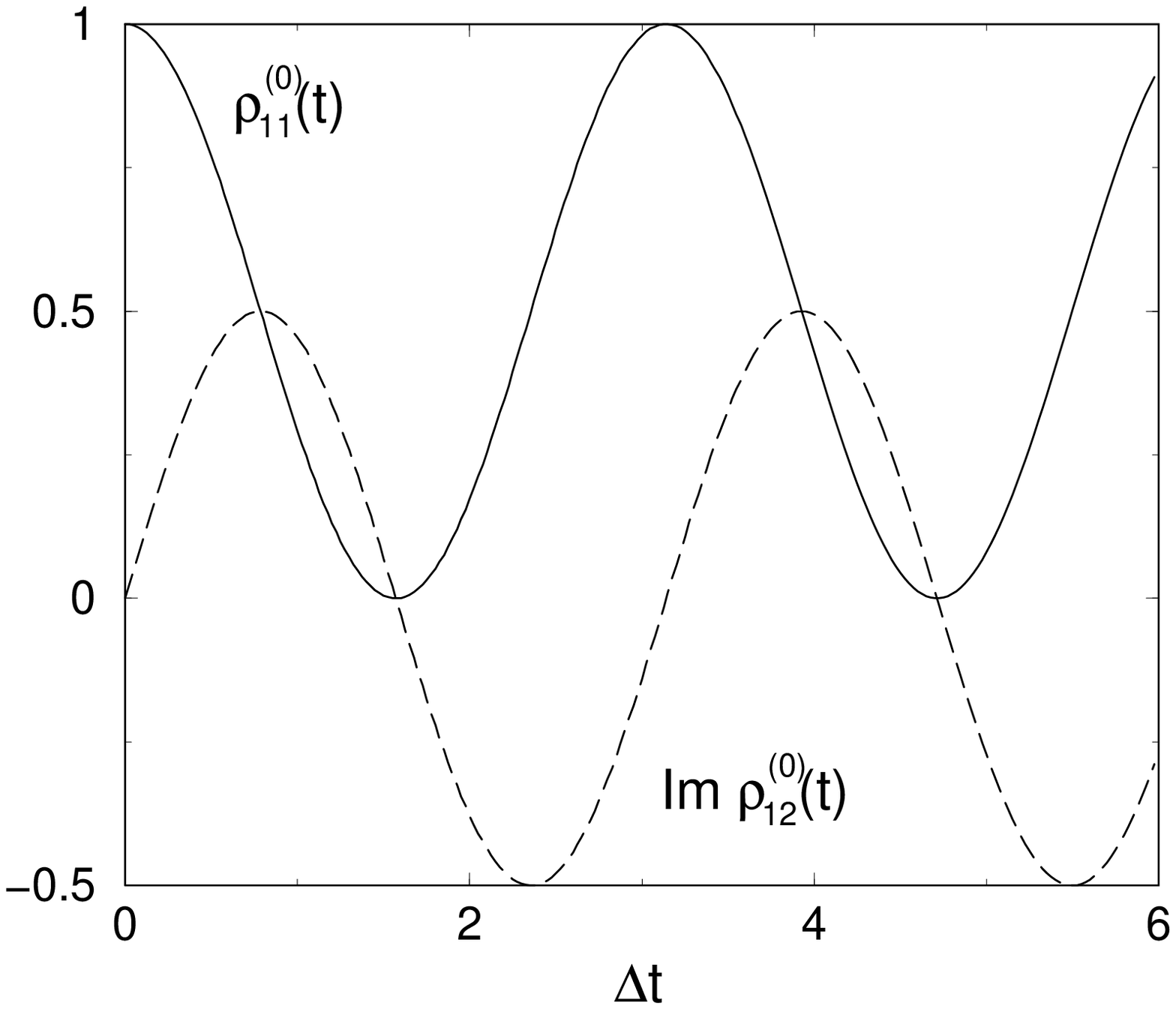}
\caption{
}
\label{fig2}
\end{figure}

\begin{figure}
\epsfxsize=12.0cm \epsfysize12.0cm
\epsfbox{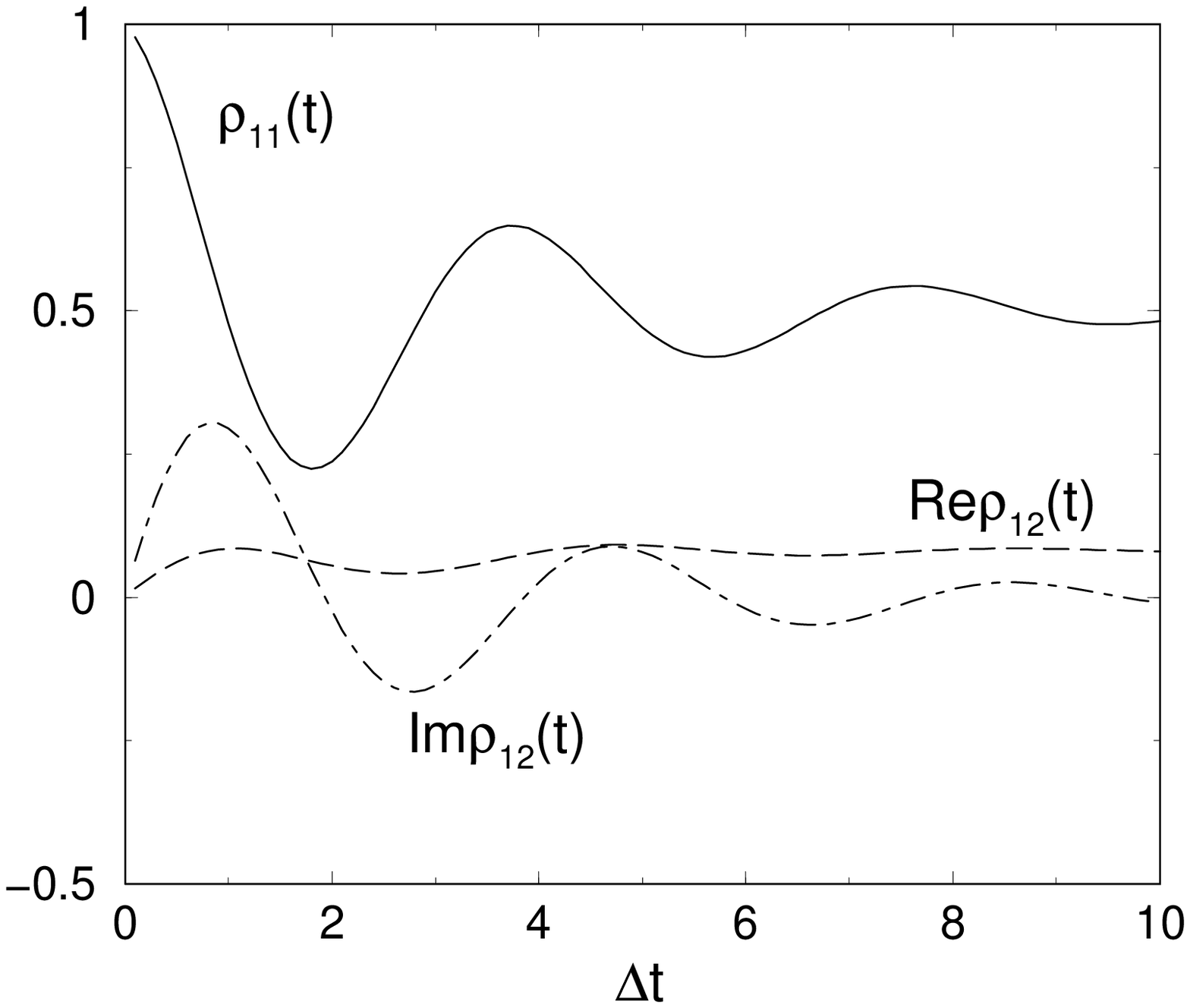}
\caption{
}
\label{fig3}
\end{figure}

\begin{figure}
\epsfxsize=12.0cm \epsfysize12.0cm
\epsfbox{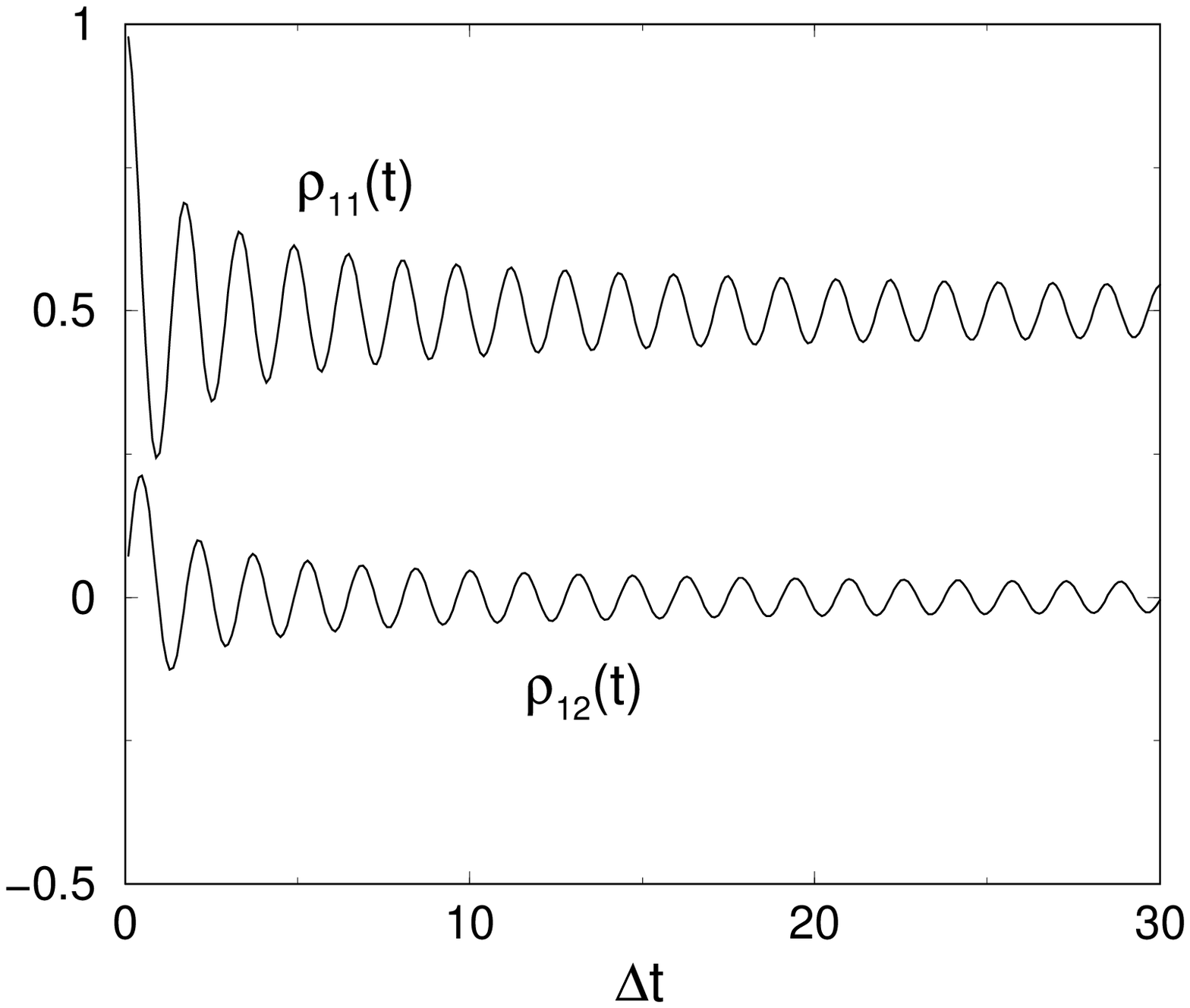}
\caption{
}
\label{fig4}
\end{figure}

\begin{figure}
\epsfxsize=12.0cm \epsfysize20.0cm
\epsfbox{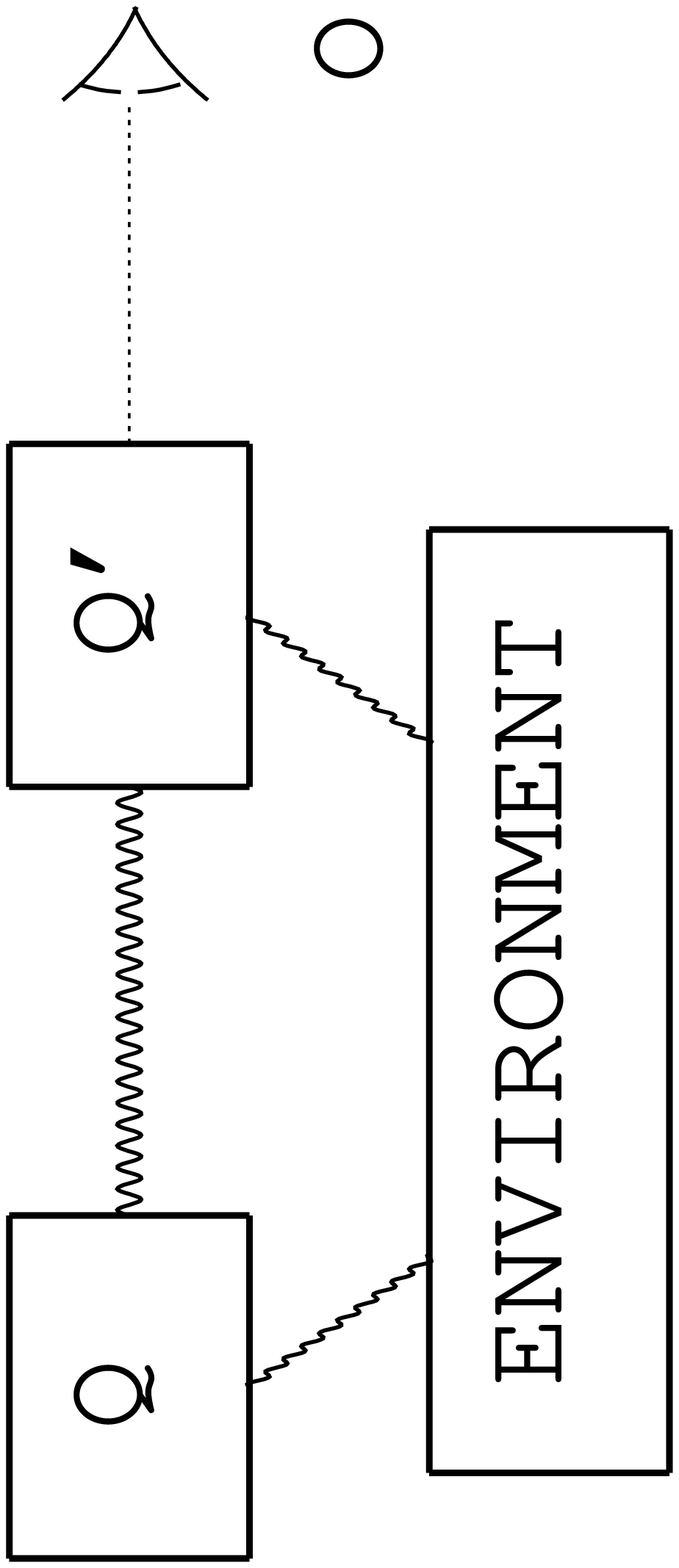}
\caption{
}
\label{fig5}
\end{figure}

Let us now consider the basis of the ``orthodox argument'' in the light of all 
this. The microscopic derivations actually do all the work for us -- 
they give an 
explicit representation of the ``microstates'' associated with each 
apparatus pointer macrostate (these being electronic, photonic, nuclear spin,
etc., states, coupled explicitly to the pointer states).
Needless to say, each macrostate is 
associated with a vast number of microstates -- however their number is not 
directly relevant to the problem. What counts is {\it how} they couple to the  
macrostate- in particular, how each individual microstate is affected by the
transition $\vert \uparrow \rangle \leftrightarrow \vert \downarrow \rangle$.
Here several crucial physical points arise:

(i) The important oscillator bath states are either gapped, by an 
energy $E_G$  typically much greater than the frequency scale 
$\omega_o$ involved in
the apparatus dynamics (eg., the Josephson plasma frequency for a SQUID);
this is the case for the electronic excitations in a SQUID or the magnons in 
a magnet. Any remaining gapless excitations (phonons, photons, etc.) have 
very little spectral weight and extremely weak coupling to the apparatus 
macro-coordinate at these frequencies (recall Fig. \ref{fig1}). 
If the electronic excitations are 
{\it not} gapped, then their effect is drastic- one gets the famous 
orthogonality catastrophe \cite{orthog,PS00} which immediately renders either 
the SQUID dynamics \cite{ajl87} or the nanomagnet dynamics \cite{PS96} 
both incoherent and strongly damped.

(ii) Gapped excitations are only very weakly disturbed by the slow apparatus 
dynamics- this is a typical example of quasi-adiabatic perturbation of the 
environmental microstates, which can be formulated to lowest approximation 
\`a la Born-Oppenheimer (and more rigourously using the effective Hamiltonian 
approach). We can think of the perturbation of the microstates as a slow
readjustment to a changing underlying vacuum state (this analogy becomes
precise when the apparatus state corresponds to that of a quantum soliton, as 
indeed it does both for SQUIDs (a tunneling fluxon) and when the magnetic 
transition corresponds to the tunneling of a magnetic domain wall \cite{sta92}.

(iii) The real danger comes from low-lying spin bath states, where this quasi-
adiabatic argument no longer applies. In the case of nuclear spins we are 
often saved by an ``anti-adiabatic'' argument- the characteristic energy 
scale of the 
spin bath states (resulting from the interaction itself) is now 
much {\it lower}
than $\omega_o$, and so again, they are hardly affected by the apparatus
transitions (the quantitative measure of the effect is provided by
the parameters ${\bf \lambda}$ 
and/or $\kappa$, whichever is applicable -- see previous
section). In the case where some of the spin bath couplings $\omega_k$ are
similar to $\omega_o$, we have the case of ``loose spins'' \cite{PS93}, where
these spins are strongly affected by an apparatus transition- the apparatus 
dynamics again becomes completely incoherent and highly damped.

From these remarks we see that the main flaw in the orthodox argument is 
simply that it never attempts a proper quantitative 
discussion of the microstates it
invokes. However, to get further insight into the physics, and to cover the 
problem of coupled qubits, we must go to the dynamics of the PISCES model. 


(iv) {\bf Dynamics}: 
The dynamics of this model in the absence of external biases 
was solved by us a number of years ago \cite{PIS}. 
It is controlled by the 
interaction ${\mathcal K}$, the tunneling matrix elements $\Delta_1$
and $\Delta_2$, the temperature $T$ and three friction coefficients
$\alpha_1$, $\alpha_2$ (describing the Ohmic coupling of each spin to the 
bath), and $\alpha_{12}$ (which describes the effective friction acting on the 
mutual correlations of the spins). 
Away from the perturbative limit ${\mathcal K} \ll \Delta$, 
the spins can be either totally locked (if ${\mathcal K} \gg T$, which 
results in a {\it single} spin-boson system with tunneling matrix 
element $\tilde{\Delta} \sim \Delta_1 \Delta_2 / {\mathcal K}$) or,  
if ${\mathcal K} \ll T$, still in a correlated phase but strongly
disordered by temperature. Coherent behaviour of the 2-spin complex is
however possible in the ``Mutual Coherence Phase''
($T \gg \Delta_j/ \alpha_j$, ${\mathcal K}$, where $j=1,2$). 
This phase then allows situations
in which the combined system-apparatus complex is in a a coherently 
entangled macroscopic state; or in the case of coupled qubits, where 
a computation can be done in time scales much shorter than the decoherence
time for this state. At this point we  
note an important feature of this model-  
decoherence effects are massively amplified as soon as
${\mathcal K} > \Delta_{\beta}$. 

Let us first use the results to set up a measurement
scheme treating both a measuring apparatus $\tau_A :
\{ \Uparrow, \Downarrow \}$
and a system $\tau_s : \{ \uparrow, \downarrow \}$ in a 
fully quantum way {\it and} in presence of an environmental bath. 
To be specific, we consider the PISCES model, with bare Hamiltonian
\begin{equation}
H_0 = \Delta_A \tau_{A}^{x} + \Delta_s \tau_{s}^{x} - \xi_A \tau_{A}^{z}
- {\mathcal K} \, \tau_{A}^{z} \tau_{s}^{z}
\end{equation}
such that ${\mathcal K} \gg \xi_A \gg \Delta_A, T$ 
represents a strong ferromagnetic interaction and
$\Delta_A \gg \Delta_s$; $\alpha_A \geq 1$. Without coupling, the apparatus
is in state $|\Uparrow \rangle$ and the restriction 
$\Delta_A \gg \Delta_s$ insures that the apparatus reacts quickly
to any changes in the system.

With the apparatus initially in the state $| \Uparrow \rangle$ and the
coupling turned on at some time $t$, then a state 
$| \Uparrow \uparrow \rangle$ remains due to bias $\xi_A$. On the other 
hand, if at time $t$ the coupled state is 
$| \Uparrow \downarrow \rangle$,
fast relaxation to the state $| \Downarrow \downarrow \rangle$ then occurs 
(at a rate 
$\tau_{A}^{-1} \sim (\Delta_A^2/\omega_0) 
(K/\omega_0)^{2\alpha_A -1}$) and the combined apparatus-system is then
essentially frozen. Relaxation to the state 
$| \Uparrow \uparrow \rangle$ takes place at the extremely slow rate
$\tau_{c}^{-1} \sim (\Delta_A \Delta_s/{\mathcal K} \omega_0) 
(T/\omega_0)^{2\alpha_A -1} \ll \tau_{A}^{-1}$. In effect, an ``ideal''
measurement of the form \cite{hepp}
\begin{eqnarray}
& & |\Uparrow \rangle |\uparrow \rangle \longrightarrow |\Uparrow \rangle 
|\uparrow \rangle  \nonumber
\\
& & |\Uparrow \rangle |\downarrow \rangle \longrightarrow |\Downarrow \rangle 
|\downarrow \rangle  \;
\end{eqnarray}
has been performed. The difference in time scales is what makes 
the measurement possible.

Now, let us look at another limit of this model, more appropriate to the 
discussion of decoherence in a pair of coupled qubits. All external biases 
are removed and we assume that (a) each system is identical, and (b) that
the dissipative couplings $\alpha = \alpha_1 = \alpha_2$ are very small.
If we wish to operate the coupled system as a pair of qubits, what then
will be the decoherence rates? let us consider the specific situation  
where the 2 qubits are in the pure state $| \up \up \rangle$ at time $t=0$ 
and calculate the diagonal elements $\rho_{\nu_1 \nu_2} (t)$ with 
$\nu_1 = \{ \up, \down \}$ and $\nu_2 = \{ \up, \down \}$. Without going into 
details (which are complex- the time evolution of the density matrix involves 
three charactristic frequencies and 3 different relaxation times 
\cite{PIS}) one can 
make the following general remarks. First, the 
general relaxation of the ``ferromagnetic''  state, 
as measured from the time behaviour of the combination 
$\rho_{\up \up} (t) + \rho_{\down \down} (t)$, is 
controlled by combinations of 
the decay rates $\Gamma_{11} ( 2 \Delta_r, {\mathcal K})$
and $\Gamma_{12} (2 \Delta_r, {\mathcal K})$, using the same notation as 
we used above for the diagonal
and off-diagonal matrix elements 
of the single spin-boson problem in Eqs. (\ref{g11})
and (\ref{g12}), and replacing $\gamma (E)$ by $\gamma 
(\sqrt{{\mathcal K}^2 + \Delta^2})$. 
On the other hand, the ``dephasing'' to the states
$| \up \down \rangle$ and $| \down \up \rangle$ occurs essentially
at rates $\Gamma_{11} (\Delta_r, {\mathcal K})$ and
$\Gamma_{12} (\Delta_r, {\mathcal K})$.
These are also the decay rates for the 
antiferromagnetic global combination $\rho_{\up \down} (t) +
\rho_{\down \up} (t)$. Thus the principal result is that the 
mutual coherence or ``entanglement'' correlations are very fragile- they are 
rendered incoherent by a combination of environmental couplings and the 
inter-qubit coupling ${\mathcal K}$ itself, in a way very much like the 
destruction of coherent oscillations for a single qubit once a bias is 
applied (note how ${\mathcal K}$ plays much the same role in the 2-qubit
entanglement relaxation as the bias $\xi$ does in the single qubit relaxation).

However we also see that if $\alpha$ is small enough 
(ie., that $\Gamma/\Delta \ll 1$) then decoherence will not be a problem 
for the operation of a pair of qubits. One can go into more details here, 
looking at the regions in the 
parameter space of variables ${\mathcal K}, T, \alpha$, and $\Delta$ where 
a pair of coupled qubits can operate (the ``mutual coherence regime'' defined
in Dub\'e and Stamp \cite{PIS}), but this discussion becomes rather technical 
so we refrain from this here. We also note that any realistic analysis
of this dynamics must also include the spin bath. 

In any case our main point is clear - by fiddling with the 
parameters of this coupled system 
we can make it go from a coherent regime to an incoherent regime. In 
the latter the apparatus really acts as a measuring device, and in the 
former as a pair of qubits. 
We note also that a crucial feature of the apparatus when in the conventional 
measurement mode is that it is sufficiently damped to  
settle into a single ``macrostate''. This is of course is hardly new- it is
follows from the very definition of a "measurement" that this must be so!
But this does not affect our point- which is that
the same system that can behave as a measuring apparatus can also, with 
only small changes in coupling constants or fields, show macroscopic
superpositions. One may try to argue (like Peres 
\cite{complex,peres}) that the word ``macroscopic'' {\it by definition}
ought to imply that macroscopic superpositions are disallowed (so that coherent 
superpositions of SQUID flux states, or magnetic 
magnetisation states, must by definition be considered microscopic no
matter how large the SQUIDs, magnets, or 
flux or magnetisation differences may be). But then 
the whole orthodox argument becomes a mere tautology, and explains nothing- 
as well as forcing a very peculiar notion of the term ``macroscopic'' upon us
(one in which a system changes from macroscopic to microscopic by, eg., 
tuning of an external field!).

We finish by stressing that these considerations are not academic. 
Experimentalists searching for large-scale quantum effects in magnets and
superconductors must push their 
SQUID magnetometers or nanomagnetic detectors into the quantum
regime to attain increased sensitivity. They will eventually 
have to take account of quantum correlations in the
functioning of the measuring apparatus. This promises to lead to all sorts of
interesting discussions! And of course a world-wide effort is on to build 
solid-state qubits. Thus the decoherence story is only just beginning.

ACKNOWLEDGEMENTS

We would like to thank FOM and NSERC Canada for support of this research,
and also Caspar van der Wal for discussion of his experimental results.


\newpage

{\bf Figure Captions}

\vspace{1cm}

Fig. 1: Typical form for the dimensionless coupling $g(\omega, T)$, with
$\omega = T$, between
a mesoscopic system such as a superconductor or a nanomagnet, and various 
environmental modes. In this particular case we assume a 3-dimensional system
with nuclear spin energy scale $E_o \sim 3 \times 10^3~K$, and a 
superconducting transition at $T \sim 2~K$. 

\vspace{1cm}

Fig. 2: Diagonal and Off-diagonal  of the density
matrix for a "Free spin", uncoupled from the environment, and in
zero bias field ($\xi =0$).

\vspace{1cm}

Fig. 3: Diagonal and Off-diagonal elements of the density
matrix in the spin-boson problem with parameter $\alpha=0.05$,
$T/\Delta = 0.5$ and $E_c = 50 \Delta$. The value
$\rho_{12} (t \rightarrow \infty) = (\Delta / \Delta_r) \tanh
( \Delta_r /2 T)$. The system is again unbiaised, ie., $\xi =0$.

\vspace{1cm}

Fig. 4: Diagonal and Off-diagonal elements of the density
matrix in the central spin problem with zero bias, and
with topological decoherence dominating.
The value of the topological decoherence 
parameter is relatively weak: $\lambda = 1/8$.

\vspace{1cm}

Fig. 5: The general problem involved in dealing with a measuring 
system (collective coordinate $Q^{\prime}$ coupled to some other 
quantum system (collective coordinate $Q$), with both coupled to a 
quantum environment; all variables are quantised. The same kind of model 
describes a pair of coupled qubits.

\newpage

\newpage

\newpage

\newpage

\newpage

\end{document}